\documentclass{article}[11]
\usepackage{graphicx}
\usepackage{epstopdf}

\begin{document}
\title{Effective Conductivity of Spiral and other Radial Symmetric Assemblages}
\author{Andrej Cherkaev \\Department of Mathematics, University of Utah \\ and 
\\Alexander D. Pruss \\ Department of Mathematics, Duke University}
\date{June 20, 2012}

\maketitle
\abstract{Assemblies of circular inclusions with spiraling laminate structure inside them are studied, such as spirals with inner inclusions, spirals with shells, assemblies of "wheels" - structures from laminates with radially dependent volume fractions, complex axisymmetric three-dimensional micro-geometries called Connected Hubs and Spiky Balls. The described assemblages model structures met in rock mechanics, biology, etc. The classical effective medium theory coupled with hierarchical homogenization is used. It is found that fields in spiral assemblages satisfy a coupled system of two second order differential equations, rather than a single differential equation; a homogeneous external field applied to the assembly is transformed into a rotated homogeneous field inside of the inclusions. The effective conductivity of the two-dimensional Star assembly is equivalent to that of Hashin-Shtrikman coated circles, but the conductivity of analogous three-dimensional Spiky Ball is different from the conductivity of coated sphere geometry.}

\section{Introduction}

Structures with explicitly computable effective properties play a special role in the theory of composites. They allow for testing, optimizing, and demonstrating of dependences on the structural parameters and material properties. These structures are also used for  hierarchical modeling of more complicated structures and they permit explicitly computing fields inside the structure and track their dependence on structural parameters. There are several known classes of such structures: the Hashin-Shtrikman coated spheres structure \cite{hs1} and Schulgasser's structures \cite{schulgaser} are probably the most investigated geometries of composites. The scheme has been generalized to multiscale multi-coated spheres (see the discussion in \cite{cherkaev-book,milton-book}), coated ellipsoids \cite{milton-benv}, and the "wheel assembly", studied in \cite{cherk11}. Another popular class is laminate structures and derivatives of them, the laminate of a rank, which exploit a multiscale scheme in which a course scale laminate is  made from smaller scale laminates in an iterative process. The limits of iterations of these structures yields to a differential scheme where an infinitesimal layer is added at each step, see \cite{cherkaev-book,milton-book}.

In the present paper, we combine the idea of multi-rank laminates and coated spheres, introducing assemblies of circular inclusions with spiraling laminate structure inside them. Namely, we study the assemblages of spirals with inner inclusions, spirals with shells, and assemblies of "wheels" - structures from laminates with radially dependent mass fractions. We also derive the effective conductivities of complex three-dimensional microgeometries, which we call \emph{Connected Hubs} and \emph{Spiky Balls}. The described structures model inhomogeneous materials met in rock mechanics, biology, etc. For calculating effective properties, we use the classical effective medium theory, see \cite{hs1,milton-book,nasser,rh57}) coupled with hierarchical homogenization. We study spiral assemblies with inclusions and observe an interesting phenomenon: a homogeneous external field applied to the assembly is transformed into a rotated homogeneous field inside of the inclusions. The fields in such structures satisfy a coupled system of two second order differential equations, rather than a single differential equation that is satisfied in Hashin-Shtrikman and Schulgasser's structures. We show that the effective conductivity of the two-dimensional Star  is equivalent \cite{cherk11} to conductivity of Hashin-Shtrikman coated circles, but the conductivity of the analogous three-dimensional Spiky Ball is different from the coated spheres geometry.
 
\section{Composite circular inclusion}

Consider an infinite conducting plane with coordinates $(x_1,x_2)$ and assume that a unit homogeneous electrical field $e=(1,0)^T$ is applied to the plane at infinity, inducing a potential $u$,
\begin{equation}
\lim_{||x||\to\infty} u(x) = x_1.
\label{bdryatinfty}
\end{equation}
A structured circular inclusion of unit radius $\|x\| \leq 1$ is inserted in a plane. It consists of a core inner circle of radius $r_0$ and an enveloping annulus. The inner circle (nucleus) $\Omega_i = \{x:||x||<r_0\leq 1$\}, is  filled with an isotropic material of conductivity $\sigma_i$. The annulus $\Omega_a = \{x: r_0<||x||<1\}$ is filled with anisotropic material whose conductivity tensor $S_e(r)$ depends only on radius. The plane outside the inclusion is denoted $\Omega_*$, it is filled with an isotropic material of conductivity $\sigma_*$. 

Effective conductivity of such an assembly is computed by effective medium theory. Given the inclusion, we find the conductivity $\sigma_*$ so that the inclusion is cloaked,
\begin{equation}
u(x) = x_1 \quad \mbox{ if } ||x||>1. \label{bdryatinfinity}
\end{equation}
If this is the case, the outer conductivity $\sigma_*$ is called the effective conductivity of the inclusion. The inclusion is not seen by an outside observer, therefore the entire plane can be filled with such inclusions, according to  effective medium theory \cite{hs1,milton-book,nasser,rh57}.

We show that the field inside the nucleus $\Omega_i$ has the representation
\begin{equation}
u = \rho\left(\cos(\psi) x_1 +\sin(\psi) x_2 \right), \quad  ||x||<r_0
\end{equation}
where $\rho, \psi$ are constant. An observer inside $\Omega_i$ records a homogeneous field similar to the outside field, but rotated by an angle $\psi$. 

The current $j$,  electric potential $u$, and  electric field $e$ in a conducting medium are related by equations
\begin{equation}
\nabla \cdot j = 0, \quad e = \nabla u, \quad
j = K\,e , \label{currentandfield} 
\end{equation}
where $K$ is a positively defined symmetric conductivity tensor that represents the material's properties. In our assemblage,
$$K = \left\{ \begin{array}{ccc}
\sigma_i\mbox{Id}&\mbox{ if }&x\in\Omega_i\\
S_e&\mbox{if}&x\in\Omega_a\\
\sigma_*\mbox{Id}&\mbox{ if }&x\in\Omega_*
\end{array}\right. , 
$$ 
Here, Id is the identity matrix. 
Equations (\ref{currentandfield}) are combined as a conventional second order conductivity equation
\begin{equation}
\nabla \cdot K \nabla u = 0. \label{mainequation}
\end{equation}

On the boundaries between different regions, tensor $K$ is discontinuous. But the potential $u$, the normal current $j \cdot n$, and the tangential potential $e \cdot t$ are continuous,
\begin{equation}
[u]_-^+=0, \quad [j \cdot n]_-^+ = 0, \quad [e \cdot t]_-^+ = 0, \label{fieldbdry}
 \end{equation}
where $[.]_-^+$ denotes the jump. For instance, at the boundary of the nucleus, we have
$$[u]_-^+ =\lim_{\epsilon \to 0, \epsilon > 0} u(r+\epsilon,\theta)-u(r-\epsilon,\theta).$$

\section{Fields in a spiral assemblage}

In this section we find the fields, currents, and effective conductivity of the described assemblage assuming that the eigenvectors of $K$ in $\Omega_a$ form a family of logarithmic spirals. We call the resulting structure the Spiral with Core.

\begin{figure}[htp]
\centering
\includegraphics[width = 3cm]{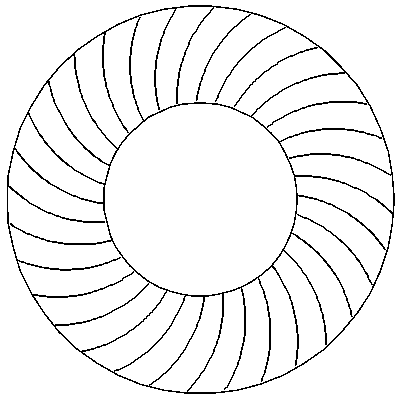}
\caption{Spiral with Core}
\end{figure}

\textbf{Single Inclusion.} Rewrite the problem in polar coordinates $r,\theta$. Solving conductivity equation (\ref{mainequation}), we find that the potential in the inner and outer isotropic regions is

\begin{eqnarray}\left\{ \begin{array}{llll}
u(r,\theta) = A\,r\cos(\theta)+B\,r\sin(\theta)& \mbox{ if }  & 0\leq r < r_0 & (x\in\Omega_i)\\
u(r,\theta) =  r\cos(\theta)&\mbox{ if } & r>1, & (x \in \Omega_*) \label{outside}
 \end{array}\right.
\end{eqnarray}
$A,B$ are constants. The form of the solution in the outer domain reflects the effective medium condition: the inclusion is invisible, and  field is unperturbed and agrees with condition (\ref{bdryatinfinity}).

 Assume that the angle $\phi$ of orientation of the principle axes of $K$ in $\Omega_a$ to the radius is constant, so that the eigendirections form a family of logarithmic spirals.  In $\Omega_a$, $K$ has the form
$$K = \left[ \begin{array}{ll}
K_{rr}&K_{r\theta}\\
K_{r\theta}&K_{\theta \theta} \end{array}\right]
= RS_e R^T,
$$
where
$$
R = \left[\begin{array}{ll}
\cos(\phi)&\sin(\phi)\\
-\sin(\phi)&\cos(\phi) \end{array}\right], \quad S_e = \left[ \begin{array}{ll}
\sigma_1&0\\
0&\sigma_2 \end{array} \right].
$$
$\sigma_1$ and $\sigma_2$ are positive constant eigenvalues of $S_e$, and 
$\phi$ is the angle of the spiral.  The entries of $K$ are: 
$$\begin{array}{rrr}
K_{rr}&=&\sigma_1\cos^2(\phi)+\sigma_2\sin^2\phi,\\
K_{r\theta}&=&(\sigma_1-\sigma_2)\cos(\phi)\sin(\phi)\\
K_{\theta \theta}&=&\sigma_1\sin^2(\phi)+\sigma_2\cos^2(\phi)\end{array}$$

Equation (\ref{mainequation}) in the annulus becomes
\begin{equation}
\left(\frac{1}{r}\frac{\partial}{\partial r}r\left[ K_{rr} \frac{\partial}{\partial r}+ K_{r\theta} \frac{1}{r}\frac{\partial}{\partial \theta}\right]+ \frac{1}{r}\frac{\partial}{\partial \theta}\left[   K_{r\theta} \frac{\partial}{\partial r}+ K_{\theta \theta} \frac{1}{r}\frac{\partial}{\partial \theta}  \right]\right) u(r, \theta)=0
\end{equation}
We separate variables and account for boundary conditions. The solution has the form 
$$u(r,\theta) = U(r)\cos(\theta) + V(r)\sin(\theta).$$
%
%
%
and $U$ and $V$ satisfy a system of ordinary differential equations
\begin{eqnarray}
L_1U - L_2 V = 0, \quad L_1V + L_2U = 0
\label{stareqn}
\end{eqnarray}
where $L_1$ and $L_2$ are linear differential operators, 
\begin{equation}L_1 = \left(\frac{d }{d r} r \, K_{r} \frac{d }{d r} - \frac{1}{r} K_{\theta \theta} \right), \quad L_2 = -\left( \frac{d }{d r} K_{r\theta}  + K_{r \theta} \frac{d  }{d r}\right).\end{equation}
Notice that equations in (\ref{stareqn}) are respectively the real and complex part of a complex-valued differential equation
\begin{equation}
(L_1+iL_2)(U+iV) = 0
\end{equation}


Write the current in the spiral as $j = J_u \cos(\theta) + J_v \sin(\theta).$ We compute
\begin{equation} 
J_u = K_{rr}\frac{dU}{dr}+K_{r\theta}\frac{V}{r}, \quad J_v = K_{rr}\frac{dV}{dr}-K_{r\theta}\frac{U}{r}.
\label{jujv}
\end{equation}
The jump conditions (\ref{fieldbdry}) at $r=r_0$ and $r=1$ have the form 
\begin{eqnarray}
 \quad [U]_-^+=0,~~ [V]_-^+ = 0,~~
[J_u]_-^+ = 0, ~~  [J_v]_-^+ = 0.
\label{currentbdry}
\end{eqnarray}

Potentials $U, V$ in annulus $\Omega_a$ are found by solving (\ref{stareqn}):
\begin{eqnarray}
U =  r^\alpha(C_1 \cos(\beta) - C_2 \sin(\beta)) + r^{-\alpha}(C_3\cos(\beta) - C_4\sin(\beta)), \label{spiralsoln1} \\
V =  r^\alpha(C_1 \sin(\beta) + C_2 \cos(\beta)) + r^{-\alpha}(C_3 \sin(\beta) + C_4 \cos(\beta)) \label{spiralsoln2},
\end{eqnarray}
where
$$\alpha = \sqrt{\sigma_1\sigma_2}, \quad \beta = \frac{\ln(r)K_{r \theta}}{K_r}.$$

\textbf{Determination of Constants.} Jump conditions (\ref{currentbdry}) at the outer boundary ($r=1$) are
\begin{equation}
C_2+C_4 = 0, \quad C_2-C_4=0\end{equation}
yielding $C_2 = 0, C_4 = 0$.  
There remain five unknown constants, $k_*, C_1, C_3, A,$ and $B$. They appear to be overdetermined by the remaining six boundary conditions. However, the four boundary conditions (\ref{currentbdry}) on the inner boundary at $r_0$ can be reduced to just three conditions, allowing the system to solved. 

The potentials $U$ and $V$ in the spiral material are linked together, and knowing either one is sufficient to determine the other. Indeed, upon substituting $C_2 = 0, C_4 = 0,$ we find
\begin{equation}
V(r) = U(r) \tan(\beta(r))\end{equation} \begin{equation}\frac{dV(r)}{dr} = \frac{dU(r)}{dr}\tan(\beta(r)) + \frac{U(r)(1+\tan^2(\beta(r)))K_{r \theta}}{r K_r}.
\end{equation}
Substituting this into (\ref{jujv}), we see that the same relationship holds for the normal components of the current,
\begin{equation}
J_v = K_{r}\frac{dU(r)}{dr}\tan(\beta(r)) + \frac{U(r)}{r}K_{r \theta}\tan^2(\beta(r)) = J_u \tan(\beta(r),
\end{equation}
so we have, in the spiral,
\begin{equation}
\frac{V(r)}{U(r)} = \frac{J_v(r)}{J_u(r)} = \tan(\beta(r)).
\label{sproof}
\end{equation}
 Using (\ref{sproof}), the current conditions (\ref{currentbdry}) are redundant, as the two conditions on the current are simultaneously satisfied. The four boundary conditions can be rewritten as three,
\begin{equation}
[U]_-^+ = 0, \quad [V]_-^+ = 0, \quad \tan(\beta(r_0))=\frac{B}{A}. 
\end{equation}

With this observation, the linear system can be solved for the remaining unknowns. Among them is the effective conductivity of the outer region, $\sigma_*$, which is treated as an unknown. Notice that the annulus with spiraling material does not perturb the outside field, and contains a homogeneous inner field that is directed in a different direction than the outer homogeneous field. 

The value of $\sigma_*$ is obtained by solving the system is the effective conductivity of the structure; it is given by an explicit formula

\begin{equation} \sigma_* = \frac{\sigma_2 \sigma_1 (r_0^\gamma -1) + \sigma_i \sqrt{\sigma_1 \sigma_2}(r_0^\gamma +1)}{\sigma_i(r_0^\gamma-1) + \sqrt{\sigma_1 \sigma_2}(r_0^\gamma + 1)}, \label{sigmastar} \end{equation}
where
$$\gamma = \frac{2 \sqrt{\sigma_1 \sigma_2}}{\sin(\phi)^2 \sigma_1 - \sin(\phi)^2 \sigma_2 - \sigma_1}.$$

\begin{figure}[htp]
\centering
\includegraphics[width = 6cm]{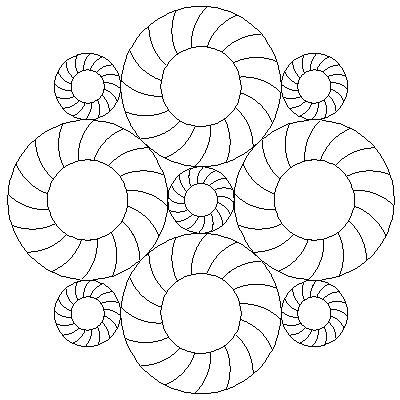}
\caption{Spiral Assemblage}
\end{figure}

\textbf{Effective Medium Theory.} The Spiral with Core structure, as described above, extends from the origin to a radius of one. But this radius is arbitrary in an infinite plane. The spiral can be scaled up or down with respect to the radius and will still solve the same problem. As the spiral leaves the outside field unperturbed, placing multiple spirals side by side will result in every object placed being rendered invisible to the field, with each having an inner homogeneous field directed in a different direction than the applied field.  While none of the myriad inclusions are detectable by perturbations of the outside field, the field is differently directed almost everywhere.


\section{Extreme Spiral Geometries}

\textbf{Extremal Spiral Angle.} By design, the core of the spiral object contains a uniform electric field directed in a different direction than the outside field. 
The potential in inner region (nucleus) has the representation 
$$u(r, \theta) = A r\cos(\theta) + B r\sin(\theta).$$
The field inside will make an angle of $\Upsilon = \tan^{-1}(B/A)$ with the uniform field outside the spiral. $A$ and $B$ depend on the spiral angle $\phi$, so a substitution shows
$\Upsilon = \gamma|\sigma_2-\sigma_1|$, where
\begin{equation}
\gamma = \frac{\ln(r_0)\cos(\phi)\sin(\phi)}{\sin^2(\phi)\sigma_1-\sin^2(\phi)\sigma_2-\sigma_1}.
\end{equation}

 Let us find the angle $\phi$ which maximizes the rotation $\Upsilon$.
\begin{figure}[htp]
\centering
\includegraphics[width = 6cm]{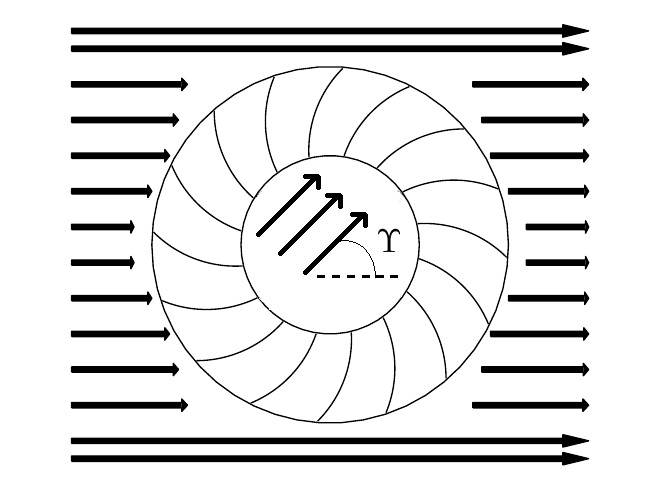}
\caption{Changing Direction of the Electric Field}
\end{figure}
A straightforward calculation shows that for a given $\sigma_1$ and $\sigma_2$,  $\Upsilon$ is maximized by choosing
\begin{equation}\phi_0 = \arctan(\sqrt{\sigma_1/ \sigma_2}).\end{equation}
The maximal angle $\Upsilon_{max}$ for a given $\sigma_1$, $\sigma_2$, and $r_0$ is
\begin{equation}\Upsilon_{max} = -\frac{1}{2} \ln(r_0) \left|\frac{\sigma_1}{\sigma_2}-1\right| \sqrt{\frac{\sigma_2}{ \sigma_1}}. \label{tmax}\end{equation}
Here, $r_0 \leq 1$, the radius of inclusion. It's clear that the more anisotropic the spiral and the smaller the inner radius $r_0$ are, the larger the resulting twist inside the spiral's core is.

\textbf{Laminates.} The conductivities $\sigma_1, \sigma_2$ in (\ref{tmax}) describe the conductivities of the outer spiral material in the Spiral with Core structure. If this anisotropic material is  a laminate made of materials with conductivities $k_1, k_2$ and volume fractions $m_1, m_2$, then $\sigma_1$, $\sigma_2$ can be written as the geometric and arithmetic means of the conductivities,\cite{cherkaev-book}

\begin{equation} \sigma_1 = m_1k_1 + m_2k_2, \quad \sigma_2 = \left(\frac{m_1}{k_1}+\frac{m_2}{k_2}\right)^{-1}.\end{equation}

For given materials with conductivities $k_1, k_2$, we can optimize $\Upsilon$ with respect to $m_1, m_2$. The maximal angle of rotation is obtained for 
\begin{equation}
m_1 = m_2 = \frac{1}{2}.
\end{equation}
It is equal to 
\begin{equation}
\Upsilon_{max} = -\frac{1}{4}\ln(r_0)\frac{(k_1-k_2)^2}{(k_1+k_2)\sqrt{k_1k_2}}
\end{equation}
In particular, $\Upsilon = \pi$ (and the current inside goes in the opposite direction) if \begin{equation}\ln(r_0) = -4\pi\frac{\sqrt{k_1k_2}(k_1+k_2)}{(k_1-k_2)^2}.\end{equation} For instance, if $k_1 = 1$ and $k_2 = 100$, the inner field will be directed opposite to the external field when $r_0 = 	0.274.$

Values of $\Upsilon$ greater than $2 \pi$ are also  possible. The relative angle between the incident current and the inner current is the value of $\Upsilon$ taken mod $2\pi$.

\section{Derivative Assemblages}

The parameters in the Spiral with Core structure can be modified to obtain effective conductivity of  similar conducting assemblages.

\textbf{Hashin-Shtrikman Coated Circles.} The classical example of the Hashin-Shtrikman geometry[3] is an isotropic circle surrounded by an isotropic annulus

\begin{figure}[htp]
\centering
\includegraphics[width = 2.5cm]{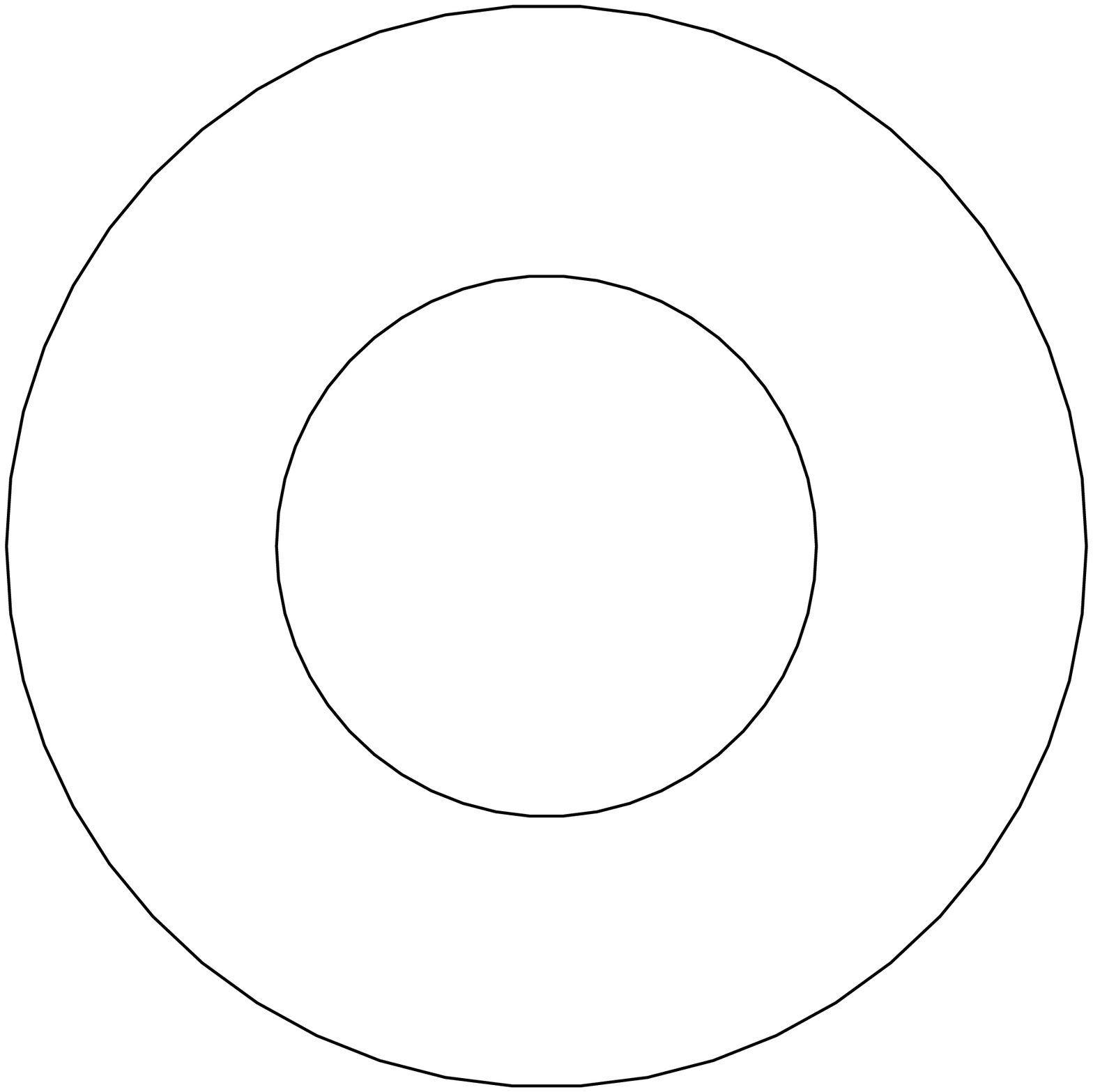} \qquad \includegraphics[width = 2.5cm]{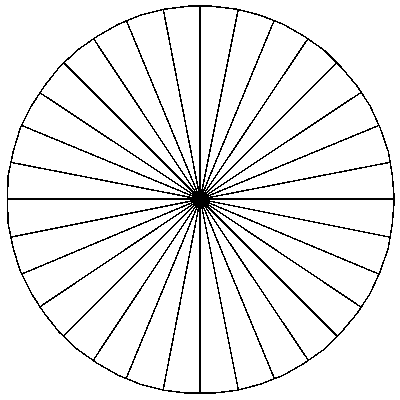}
\caption{Left: Coated Circles. Right: Schulgasser's Geometry (orange slice)}
\end{figure}
It is obtained from the Spiral with Core by setting $\sigma_1 = \sigma_2$. The outer spiral layer becomes an isotropic shell, and the effective conductivity coincides with Hashin-Shtrikman's result
\begin{equation}
k_{hs} = \sigma_1 \frac{(\sigma_i + \sigma_1) + r_0^2(\sigma_i -  \sigma_1)}{(\sigma_i + \sigma_1) - r_0^2(\sigma_i - \sigma_1)}.
\end{equation}
Notice that $m = r_0^2$ is the volume fraction of the core material.

\textbf{Schulgasser Structure.} Schulgasser \cite{schulgaser} suggested another classical symmetric geometry . It is a radial laminate of two materials.
Schulgasser's structure is obtained from the Spiral with Core by setting $r_0 = 0, \phi = 0$. The inner isotropic circle disappears and the logarithmic spiral laminate is straightened into a radial laminate.
The effective conductivity agrees with \cite{schulgaser}.
\begin{equation}
k_{sch} = \sqrt{\sigma_1 \sigma_2}.
\end{equation}

\textbf{Orange with Core.} 
This object has an inner circle of isotropic material surrounded by an annulus made up of a radial laminate.  

\begin{figure}[htp]
\centering
\includegraphics[width = 2.5cm]{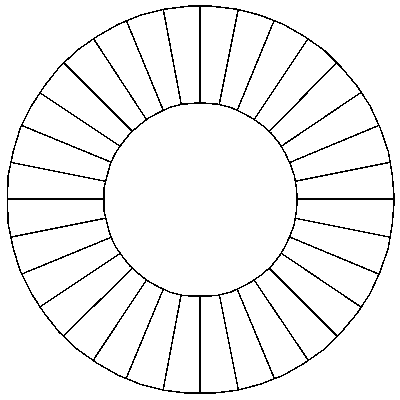}  \qquad \includegraphics[width = 2.5cm]{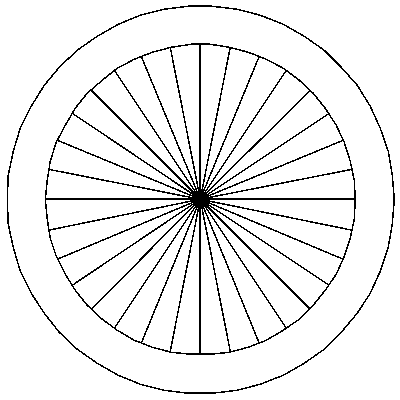}
\caption{Left Orange with Core. Right: Orange with Shell}
\end{figure}

The Orange with Core is obtained from the Spiral with Core by setting $\phi = 0$. The logarithmic spiral laminate is straightened into a radial laminate.
The effective conductivity of the Orange with Core is 


\begin{equation}
k_{c} = \sigma_1 \frac{\sigma_i \kappa \left(1+r_0^{2\kappa}\right)+
\sigma_2 \left(1-r_0^{2\kappa}\right)}
{\sigma_1 \kappa \left(1+r_0^{2\kappa}\right)+
\sigma_i \left(1-r_0^{2\kappa}\right)},
\quad \kappa = \sqrt{\frac{\sigma_2}{\sigma_1}}.
\end{equation}

Note that this formula can be written in terms of volume fractions by observing that  the volume fraction $m_i$ of the inner material is $ m_i=r_0^2 $.

\textbf{Orange with Shell.} This object contains an inner radial laminate surrounded by an isotropic material.
Denote the conductivity of the outer isotropic shell by $\sigma_1$, and  the conductivity tensor (in polar coordinates) for the inner laminate by $$K = 
\left[\begin{array}{rr}
\sigma_r & 0\\
0 & \sigma_\theta
\end{array}\right]$$
The effective conductivity can be found by a straight calculation. It is

\begin{equation}k_s = \sigma_1 \frac{\sqrt{ \sigma_\theta \sigma_r}(r_0^2 +1) + \sigma_1(1-r_0^2)}{\sqrt{ \sigma_\theta \sigma_r}(1-r_0^2) + \sigma_1 (r_0^2 + 1) }. \label{ks}\end{equation}
Indeed, the material in an inner circle is a Schulgasser radial laminate, which has effective property $k_* = \sqrt{\sigma_r \sigma_\theta}$. Thus, this inner material can be treated as an isotropic material with conductivity $\sqrt{\sigma_r \sigma_\theta}$. The effective property of the Orange with Shell can be obtained by substituting $\sigma_i = \sqrt{\sigma_r \sigma_\theta}$ into the equation for the effective property of Hashin-Shtrikman coated circles.

\textbf{Basic Spiral.} This object is simply the spiral material centered around the origin. 

\begin{figure}[htp]
\centering
\includegraphics[width = 2.5cm]{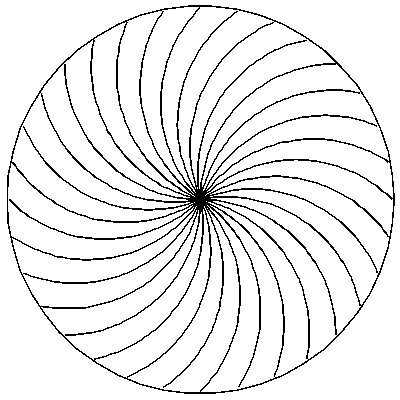} \qquad \includegraphics[width = 2.5cm]{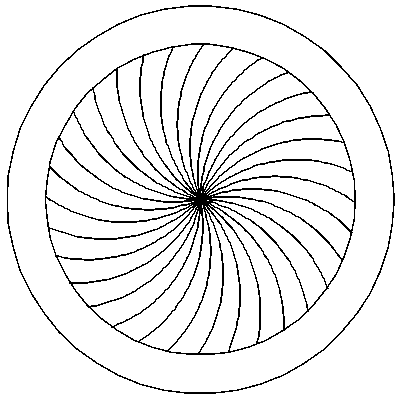}
\caption{Left: Basic Spiral. Right: Spiral with Shell}
\end{figure}

The Basic Spiral can be obtained from the Spiral with Core by setting $r_0 = 0$. 
The effective conductivity of the spiral is
\begin{equation}
k_{sp} = \sqrt{\sigma_1 \sigma_2}.\end{equation}
As one might expect, it coincides with the effective conductivity of Schulgasser's geometry.

\textbf{Spirals with Shell and Core.} The Spiral with Shell contains a circle of spiral laminate surrounded by an isotropic shell.
As we mentioned, no more work is needed to be done to calculate the effective property of it. Since the effective property of the Basic Spiral is the same as the effective property of Schulgasser's Structure, the effective property of the Spiral with Shell is the same as the effective property of the Orange with Shell, $k_s$. (\ref{ks})
For completeness, we also mention here again the Spiral with Core discussed in Section 3
that is an isotropic circle surrounded by a spiral laminate. It is a field rotator, and its effective property is given by (\ref{sigmastar}).

\textbf{Insulated Geometries.} 
Consider the asymptotic case: assume that the spiral is a laminate made from two isotropic materials, and the conductivity of one material approaches zero.





If one of the conducting materials is replaced by an insulator, the conductivities of both Schulgasser structure and the spiral are zero. The spiral with shell structure can be insulated by replacing one of the materials in the spiral laminate with an insulator. The resulting structure should behave as an annulus; the inner spiral acts as an isotropic insulator, and all current should passes through the isotropic shell. Indeed, the resulting effective property is 
$$k^0_{hs} = \lim_{\sigma_\theta \to 0} k_{hs} = \frac{1-r_0^2}{1+r_0^2} \sigma_i .
$$ 
This structure has the effective property of an annulus with insulated nucleus, as expected.

If the conductivity of one of the materials in the spiral laminate is zero, the spiral with core structure has effective property (see (\ref{sigmastar}))
$$k^0_* = \lim_{\sigma_2 \to 0} k_* =\frac{\sigma_i \sigma_r \cos^2(\phi)}{\sigma_r \cos^2(\phi) - \sigma_i \ln(r_0)}.$$
The effective property of if the insulated orange with a core obtained from this one by setting $\phi = 0$,
$$k^0_{c} = \frac{\sigma_i\sigma_r}{\sigma_r - \sigma_i \ln(r_0)}.$$

\textbf{Wheel.} 
Insulators can be used in laminates to control the variation of the current material in the layers. 
Consider the following  Wheel geometry. A central core is surrounded by an annulus that consists of conducting radial spikes surrounded by an insulator. The thickness of an individual spike is constant, and the volume fraction of conducting material in the annulus decreases with radius, because the circumference length linearly increases.
The conductivity tensor is 
\begin{equation} K = \left[ \begin{array}{ll} 
\sigma_1/r&0\\
0&0
\end{array} \right].\end{equation} A spike conducts in the radial direction but does not conduct in the circumferential direction. Call this material a spoke material.
To relate $\sigma_1$ to the conductivity $\sigma$ of the spoke material, we can write $\sigma \mu r_0 = \sigma_1$. The parameter $\mu \in [0,1]$ shows the relative thickness of the spikes at the radius $r_0$. 

This time we are dealing with a material with anisotropic conductivity that varies with radius. The solution to the potential inside the spoke material satisfies (\ref{mainequation}). The current is constant inside each spoke, therefore the potential is

\begin{equation}
u = (Ar + B)\cos(\theta) + (Cr + D)\sin(\theta).
\end{equation}

The wheel inclusion has an inner isotropic core and an outer isotropic shell, connected by spokes.
\begin{figure}[htp]
\centering
\includegraphics[width = 2.5cm]{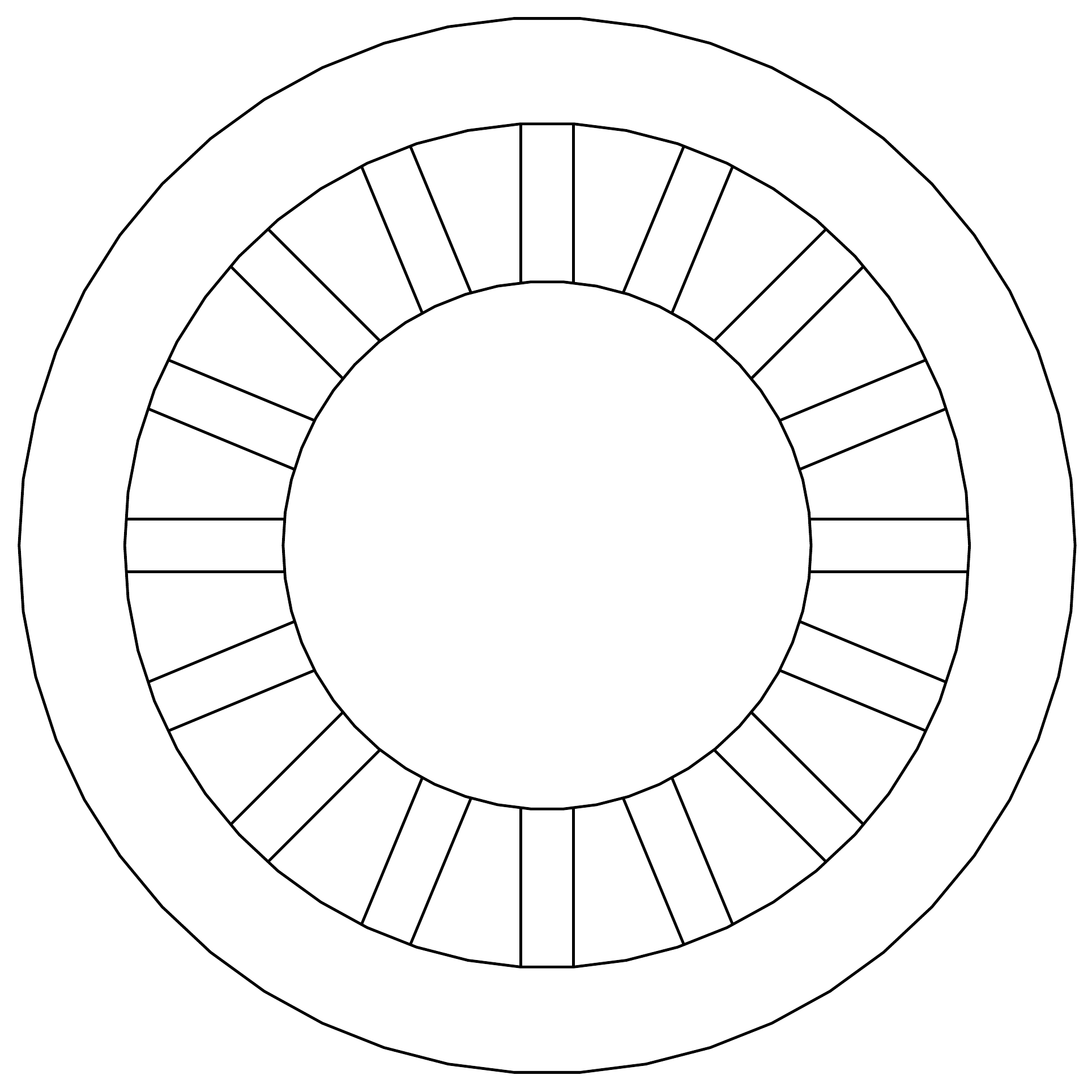}\qquad \includegraphics[width = 2.5cm]{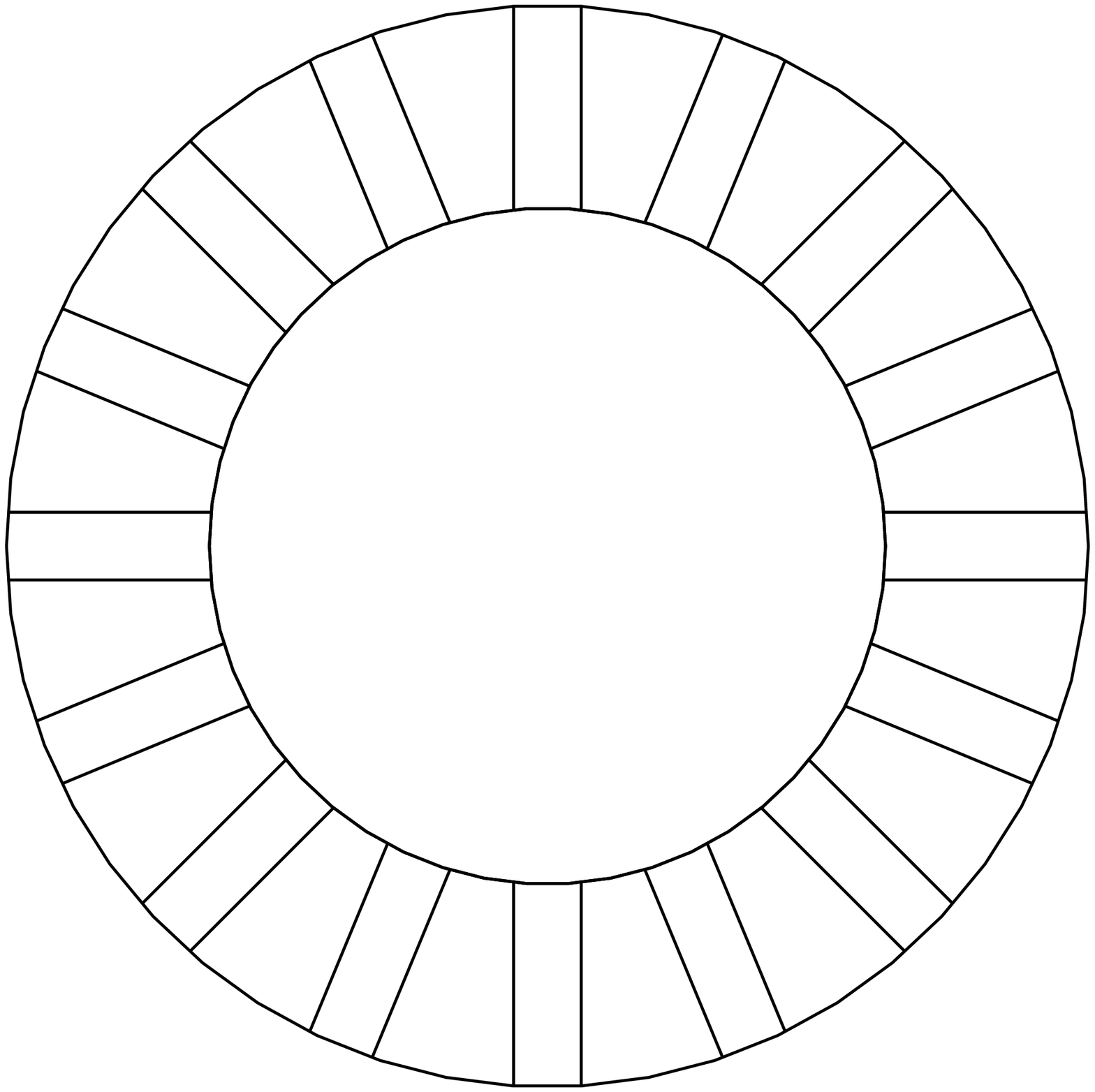}
\caption{Left: Wheel. Right: Star}
\end{figure}
It conductivity depends on radius as follows

\begin{equation}
K = \left\{
\begin{array}{lll}
\sigma_i Id& \mbox{if } & r < r_0\\
\left[
\begin{array}{ll}
\sigma_1/r&0\\
0&0\\
\end{array}
\right] & \mbox{if } & r_0< r < r_1\\
\sigma_2 Id & \mbox{if } & r_1 < r < 1\\
 \sigma_* Id & \mbox{if } & r > 1\\
\end{array}
\right. .
\end{equation}

Since the potentials in each region are known, a procedure similar to that in Section 3 gives that the effective conductivity of the Wheel:
\begin{equation}k_{wheel} = \sigma_2\frac{A+B}{B-A},\end{equation}
where

$$A = r_1^2\sigma_1\sigma_i+r_1^2r_0\sigma_2\sigma_i-r_1^2\sigma_1\sigma_2-r_1^3\sigma_2\sigma_i,$$
$$B =r_i\sigma_i\sigma_2+\sigma_2\sigma_1+\sigma_1\sigma_i-r_0\sigma_2\sigma_i. $$

 When $r_1 \to 1$, the outer conducting layer disappears and we come to the structure that we call Star.  Its effective conductivity is

\begin{equation} k_{st} = \frac{\sigma_i \sigma_1}{  \sigma_1 + \sigma_i (1-r_0)}.\end{equation}
Star assemblages can model structures of hubs connected by conducting strands in an insulating space. If the conductor in inner circles and spoke is the same, $\sigma_i = \sigma $, we write the effective property $\sigma_{star}$ in terms of the volume fraction $m$ of conducting material $\sigma$, 
\begin{equation} k_{star}(r_0,m) = \frac{\sigma m(r_0)}{2-m(r_0)}, \quad \mbox{where} \quad m(r_0) = r_0^2+2 \mu r_0(1-r_0).\end{equation}

If $\mu = \frac{1}{2}$, then $r_0 = m$ and $k_{star}$ coincides with the effective conductivity of coated circles, therefore it is an optimal structure for the resistivity minimization problem \cite{cherk11} along with the coated circles \cite{hs1}. 
The value $\mu = \frac{1}{2}$ makes intuitive sense. One can see that if the spokes in the Star cover only half the circumference of the inner sphere at $r_0$, then the current density in the isotropic region of the star will be half that in the spokes region of the star. If two orthogonal currents are separately applied to the Star structure, the sum of their energy density is constant throughout the structure. This satisfies a necessary and sufficient condition for minimization of resistivity \cite{cherkaev-book}

\section{3D radially symmetric assemblages}

The same approach can be applied to three-dimensional radially symmetric inclusions. 
Let us consider spherical inclusions, embedded in an infinite three-dimensional isotropic conducting material with undetermined conductivity. A uniform electric field is applied to the plane. The equation (\ref{mainequation})
is used to solve for the potential in each material. The boundary conditions (\ref{fieldbdry}) are similar to the two-dimensional case.  
Applying a uniform current along the $z$-axis results in the electric potential
\begin{equation}
u = \rho \cos (\phi) \end{equation}
outside the inclusion. This solution is obtained by separation of variables in equation (\ref{mainequation}) in spherical coordinates. 


\textbf{Hub.}  We construct an analogue of the two-dimensional spoke material, which consists of radial spokes of constant thickness. The relative proportion of conducting material is inversely proportional to the square of the radius, and the tensor in spherical coordinates is 
\begin{equation}
K = \frac{\sigma \mu \rho_0^2}{\rho^2}
\left[
\begin{array}{lll}
1&0&0\\
0&0&0\\
0&0&0\\
\end{array}
\right]. \label{3dspoke}
\end{equation}   We choose $\mu \in [0,1]$ so that the term $\mu \rho_0^2$ is the surface area covered by the spikes at the inner radius $\rho_0$.

The solution for the potential inside this material is
\begin{equation}
u = (C_1 + C_2\rho)\cos(\phi). \label{3dspokefield}
\end{equation}
Using this material, we construct a Hub structure. The Hub is the three-dimensional analog of the two-dimensional Star structure. It model inclusions spread through an insulating space and connected by conducting strands.
The Hub consists of an inner isotropic core of conductivity $\sigma_i$ out to a radius of $\rho$, enveloped by a spoke material. \begin{figure}[htp]
\centering
\includegraphics[width = 3cm]{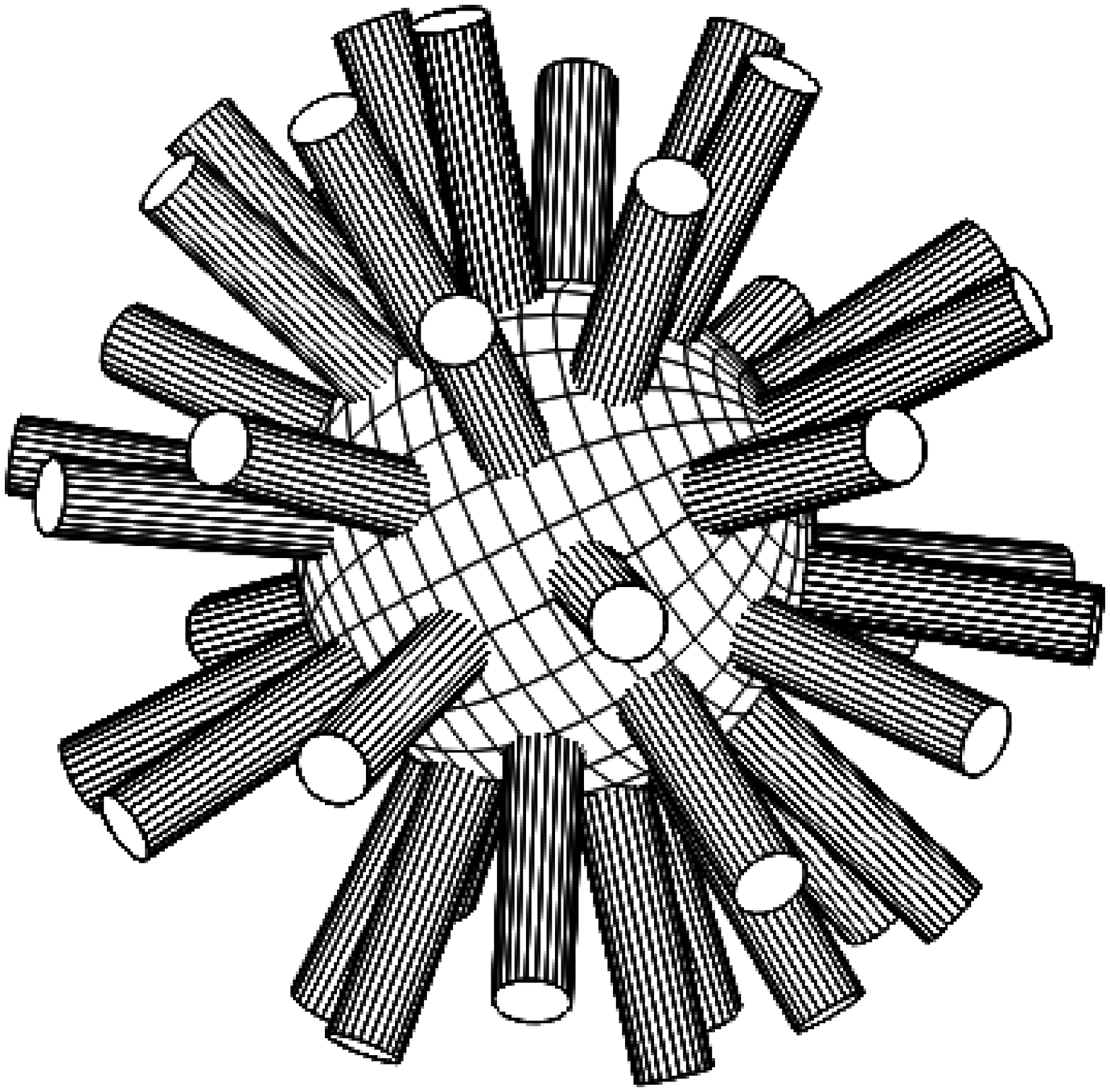} \qquad \includegraphics[width = 3cm]{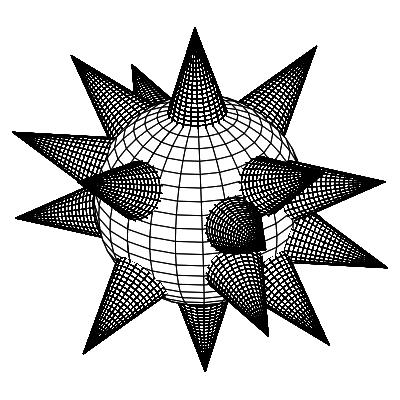}
\caption{Left: Hub. Right: Spiky Ball.}
\end{figure}

The effective conductivity of the Hub is computed similarly to that of Star. It depends on two structural  parameters, $\rho_0$ and $\mu$ and  is equal to
\begin{equation}k_{hub} = \frac{\sigma_i \sigma \rho^2_0 \mu}{\sigma_i(1-\rho_0)+\sigma \mu \rho_0}.\end{equation}

 The volume fraction of the Hub is

\begin{equation} m = r_0^3+3\mu r_0^2(1-r_0).\end{equation}
If, by analogy with the star, we take $\mu = \frac{1}{3}$, this relation gives $r_0^2 = m$, and the effective conductivity is
\begin{equation}k_{hub} = \frac{\sigma m}{3-2\sqrt{m}}.\end{equation}
Unlike the two-dimensional case, this conductivity is different from the conductivity of coated spheres.

\textbf{Spiky Ball.} The Spiky Ball structure is a generalized hub, in which the spikes have variable cross-section. The tensor of this generalized material is 

\begin{equation} K = \frac{\sigma \mu \rho_0^n}{\rho^n}
\left[
\begin{array}{lll}
1&0&0\\
0&0&0\\
0&0&0\\
\end{array}
\right].\end{equation}
The effective conductivity of this assemblage is
\begin{equation}k_{sb} = \frac{\sigma_i \sigma \rho_0^n \mu(n-1)}{\sigma_i(1-\rho_0^{n-1})+\sigma\mu\rho_0^{n-1}(n-1)}.\end{equation} 

Notice that the Spiky Ball, Hub, and Star require that the current passes sequentially through insulated spokes and a homogeneous central circle (sphere). Because of this, their effective conductivity are of harmonic mean type,
\begin{equation} \frac{1}{k_*}=\frac{a}{\sigma} + \frac{b}{\sigma_i},\end{equation}
where $a$ and $b$ are constants depending only on the geometry.

\section{Conclusion}

In constructing the Spiral with Core structure, we constructed a field rotator, e.g. a conducting structure that rotates an incident uniform electric field while remaining undetectable. Optimization results show how to maximize the angle of rotation induced by this field rotator. By modifying the Spiral with Core, a large number of other conducting structures can be constructed and their effective conductivities computed. The analogous 3D Hub and 2D Star structures behave differently in the sense that the 2D star corresponds to minimal effective conductivity while the 3D Hub does not.

~

\noindent
{\bf Acknowledgement} The work was supported by NSF through grant DMS 0707974.


\begin{thebibliography}{99}

\bibitem{cherk11} A. Cherkaev. Optimal Three-Material Wheel Assemblage of Conducting and
Elastic Composites arXiv:1105.4302 [math-ph] 22 May 2011

\bibitem{acn} N. Albin, A. Cherkaev, V. Nesi. Multiphase laminates of extremal effective conductivity in two dimensions. Journal of the Mechanics and Physics of Solids, Volume 51, Issue 10, October 2003, Pages 1773-1813.

\bibitem{milton-benv}   Y. Benveniste and G.W. Milton, New Exact Results for the Effective Electric, Elastic, Piezoelectric and other Properties of Composite Ellipsoid Assemblages, J. Mech. Phys. Solids, 51, 1773-1813, 2003.

\bibitem{cherkaev-book}  A. Cherkaev. Variational methods for structural optimization, Chapter 2. Springer Verlag NY 2000

\bibitem{christ} R. M. Christensen, Mechanics of Composite Materials. Wiley NY 1979

\bibitem{hs1}   Z. Hashin, S. Shtrikman, A Variational Approach to the Theory of the Effective Magnetic Permeability of Multiphase Materials. Applied Physics 33, 3125 (1962).

\bibitem{lipin} P. Lipinskia, El H. Barhadia, M. Cherkaouib. Micromechanical modelling of an arbitrary ellipsoidal multi-coated inclusion. Philosophical Magazine, Volume 86, Issue 10, 2006.

\bibitem{milton-book}   G. W. Milton, The Theory of Composites. Campridge University Press, 2001.

\bibitem{nasser} S. Nemat-Nasser, M. Hori.  Micromechanics: Overall Properties of Heterogenous Materials. Elsevire, NY 1999

\bibitem{rh57} J. A. Reynolds and J. M. Hough.
Formulae for Dielectric Constant of Mixtures
1957 Proc. Phys. Soc. B 70 769 

\bibitem{schulgaser}   K. Schulgasser. Sphere assemblage model for polycrystals and symmetric materials, Journal of Applied Physics, 54, 3, 1380-1382, (1983).

\bibitem{torq} S. Torquato. Effective stiffness tensor of composite media-I. Exact series expansions. Journal of the Mechanics and Physics of Solids, Volume 45, Issue 9, September 1997, Pages 1421-1448

\end{thebibliography}
\end{document}